\documentclass[runningheads]{llncs}
\usepackage{graphicx}
\usepackage{amsmath}
\usepackage{amsfonts}
\usepackage{amssymb}
\usepackage{graphicx}
\usepackage{caption}
\usepackage{subcaption}
\begin{document}
\title{A low cost small reconfigurable autonomous vehicle for research, testing and teaching of guidance, control and positioning systems}
%
%
\author{Walter Grotz \and
Santiago Moreno Gelabert \and
Emanuel Trabes \and
Carlos Federico Sosa P\'{a}ez \and
Julio Daniel Dondo Gazzano}
\authorrunning{Grotz et al.}
%
\institute{Departamento de Electronica\\
Universidad Nacional de San Luis, Argentina}
\maketitle              
\begin{abstract}

We present a small autonomous vehicle, with the purpose of aiding teachers and researchers to easily deploy different approaches to guidance, control and positioning systems, without the hassle to develop the entire modules required for such endeavor. The design was mainly guided to accomplish a very small cost, along with easy accessibility of the parts used. One of the main problems of using low cost parts is the difficulty of modeling its dynamics, mainly because they lack proper documentation and also because the time varying nature of its characteristics, due to degradation caused by usage. At the same time, any change made to the physical design of the vehicle must be reflected in the mathematical model. We develop a methodology for easy reconfiguration of the vehicle, being able to obtain detailed dynamic mathematical models of the system, as well as a the statistical models needed when implementing stochastic methods such as Kalman Filtering. Baseline example implementations of guidance, control and positioning systems are provided. The hole system is autonomous and self-contained and was tested in a low cost embedded system. 

\keywords{Autonomous Vehicle \and State Estimation \and Parameter Estimation.}
\end{abstract}
\section{Introduction}
\label{sec:introduction}

The exploration of unknown and dangerous environments is of great importance, whether it is economic (such as in the failure detection of subaquatic ducts), scientific (such as the exploration of the surface of Mars) or political (such as border patrol).

These tasks often cannot be easily performed by operators. Environments that are dangerous to human life (such as in the depths of the oceans), or where it is costly, difficult or impossible to send an operator (such as on extra-terrestrial missions), are examples of these difficulties. It is therefore important to explore alternative methods for carrying out such tasks.

Autonomous robotic systems are an extremely interesting option. They could operate for long periods of time, work in environments dangerous to human life, generate detailed real-time status reports, etc. 

Recently, great progress has been made in the implementation and use of this type of autonomous vehicle. Examples include the various probes used for the exploration of Mars, and the autonomous navigation systems developed by companies such as TESLA and GOOGLE.



Broadly speaking, the system can be divided blocks, each in charge of completing a sub-tasks.
The guide block is responsible for dictating the path the vehicle must take to complete the mission. This trajectory will be both the position $\eta_{ref}$ and the speed $\eta_{ref}$ for each instant of time. The literature on this type of system is extensive, one beeing the type of systems called \emph{Teach And Repeat}. This type of system consists of two stages. First, the autonomous vehicle is taught a certain trajectory, either by guiding it with some reference, or by controlling the vehicle remotely. Then there is the repetition phase, where the vehicle must travel the previously taught path repeatedly. Examples of these systems are those developed in works like \cite{Furgale_and_Barfoot_2010} \cite{Gao_et_al_2019}.

The positioning block estimates the location of the vehicle. It uses the measurements made by the on-board sensors to estimate the state of the vehicle, mainly the position $\hat{\eta}$ and speed $\hat{\dot{\eta}}$. The literature is also extensive on these systems. There are both odometry systems \cite{Fauser_et_al_2017}, which intrinsically present an undetermined drift in $\hat{\eta}$ throughout the mission, as well as Simultaneous Localization and Mapping (SLAM) systems \cite{Kim_2015}, which attempt to limit such drift using information obtained by visiting a previously traveled position.

Finally, the controller block uses the error $e$ between the course determined by the guidance system $\eta_{ref}$ $\dot{ref}$ and the estimate $\hat{\eta}$ $\hat{\dot{\eta}}$ to generate the control actions $u$. This finally feeds the vehicle's actuators, causing the vehicle to perform the movements necessary to complete the main task. Different types of controllers can be found in the literature, such as classic PID controllers \cite{Ma_et_al_2011}, robust controllers \cite{Hou_and_Liu_2011}, optimal controllers \cite{Junwei_and_Xiaolin_2010}, adaptive controllers \cite{HongJian_et_al_2012}, etc.

Clearly, for the vehicle to complete the main task, it needs to have an implementation for each of the sub-tasks. Thus, the researcher/teacher who wishes to design a particular solution for one of the sub-tasks requires an a-priori solution for the other tasks. It will be necessary to research, implement and fine-tune each of its implementations.

Simultaneously, educational systems that are currently on the market, such as the equipment developed by Quanser, are considerably expensive. These, in turn, are mainly closed systems, not designed to allow the user to add functionality, but to be used as designed. Thus, these systems are not available to the general public, and it is common for university laboratories in Latin America to acquire only a few quantities of these systems, always far below the number of students they have. Designing one's own system from scratch requires dedication, effort and a considerable amount of time, resources that are not always available. There is also a problem related to the use of low-cost devices, since the related documentation may be scarce, the precision of them is deficient and thus the implementation of systems that use them must be carried out very carefully to take into account these deficiencies.

Furthermore, implementations such as optimal controllers or Kalman filters require a detailed model of the vehicle's time response, which can be difficult to obtain. While a detailed model can be developed in advance, the situation where the user needs to make changes to the physical structure of the vehicle is considered. It is then necessary to have a simple methodology to obtain the updated physical parameters. It must also be taken into account that in low-cost devices the dynamics are variable in time, due to the degradation that the components suffer with use.

In this work, a small low-cost vehicle is designed, which has both the necessary basic sensors and the calculation power required for the correct implementation of a complete autonomous system. All the basic algorithms necessary to perform autonomous missions were implemented, taking into account to carefully perform robust implementations, in order to avoid the low quality of the sensors. In this way this work allows the researcher/teacher to obtain a low cost vehicle, easily accessible, which in turn allows him to concentrate on the algorithms of interest without the need to design the rest of the systems required for the autonomous operation of the vehicle. At the same time, a novel system of parameter identification is made, so that the user can easily obtain both kinematic, dynamic and probabilistic models of the vehicle. This facilitates the modification of the vehicle to the requirements of the researcher/teacher and solves the difficulty of using low-cost sensors that degrade over time. The authors believe that this platform is very useful for teaching and research of the mentioned algorithms, being a substantial contribution for research in small autonomous vehicles.

\section{Vehicle Design}
\label{sec:disenio_vehiculo}


When choosing the main components of the vehicle, different parameters were taken into account. On one hand, the sensors must be widely available on the market, so that it is easy for the researcher/teacher to acquire them. On the other hand, their cost is taken into account in order to make them as cheap as possible.

\subsection{Main Frame}

A circular base platform was chosen as the main frame, where the wheels are arranged radially in relation to the epicentre of the platform. In this way, the center of mass and the axis of movement are located at the same point within the platform, thus simplifying vehicle modeling. Thus, it is possible to make pure turns with respect to its center of mass, which makes the dynamics of movement and rotation uncoupled in the equations of state. The main frame used is the Arduino 2WD \cite{chasis}.

\subsection{Sensors}

As an intrareceptive sensor, which allows measurements of the internal state of the system, an IMU \emph{Inertial Measurement Unit} MPU 6050 was incorporated into the vehicle, which has an accelerometer and a gyroscope, thus having 6 degrees of freedom in the measurements. This sensor is very low cost and widely accessible. On the other hand, the measurements they make are not directly from position, as required in autonomous robotic applications. Therefore, the use of this sensor requires a careful study of its response, in order to counteract the low quality of the measurements. Although the data sheet is available, its low quality together with the normal degradation of the physical components, makes it desirable to make new measurements of the characteristics of this sensor, in order to obtain precise physical parameters.


\subsection{Motors and Actuators}

The motors and actuators used in this work are the following \cite{MotorYActuador}. The low cost of these modules may cause the torque of the motors to be non-symmetrical. Also, bad wheel grip may cause traction to be lost, causing the torque to be non-symmetrical as well. Furthermore, there may be both linear and quadratic friction coefficients present. All these phenomena are taken into account in the design and modeling of the system.

\subsection{Computer system}

An embedded system composed of an Arduino Uno was used to operate the motors. This allows the implementation of the PWM control signals required to correctly drive them.

At the same time, a system composed of a Raspberry pi 3 board was added as the main computer. It has the size of a credit card, while containing a 4-core ARM processor, 1 gb of ram, and an embedded video board, which can be used for signal processing in the form of images or others.

Both onboard computers are both low cost and widely available, and provide a good balance between cost and computing power.

\subsection{Complete Vehicle}

\begin{figure}[h]
    \centering
    \includegraphics[width=0.7\linewidth]{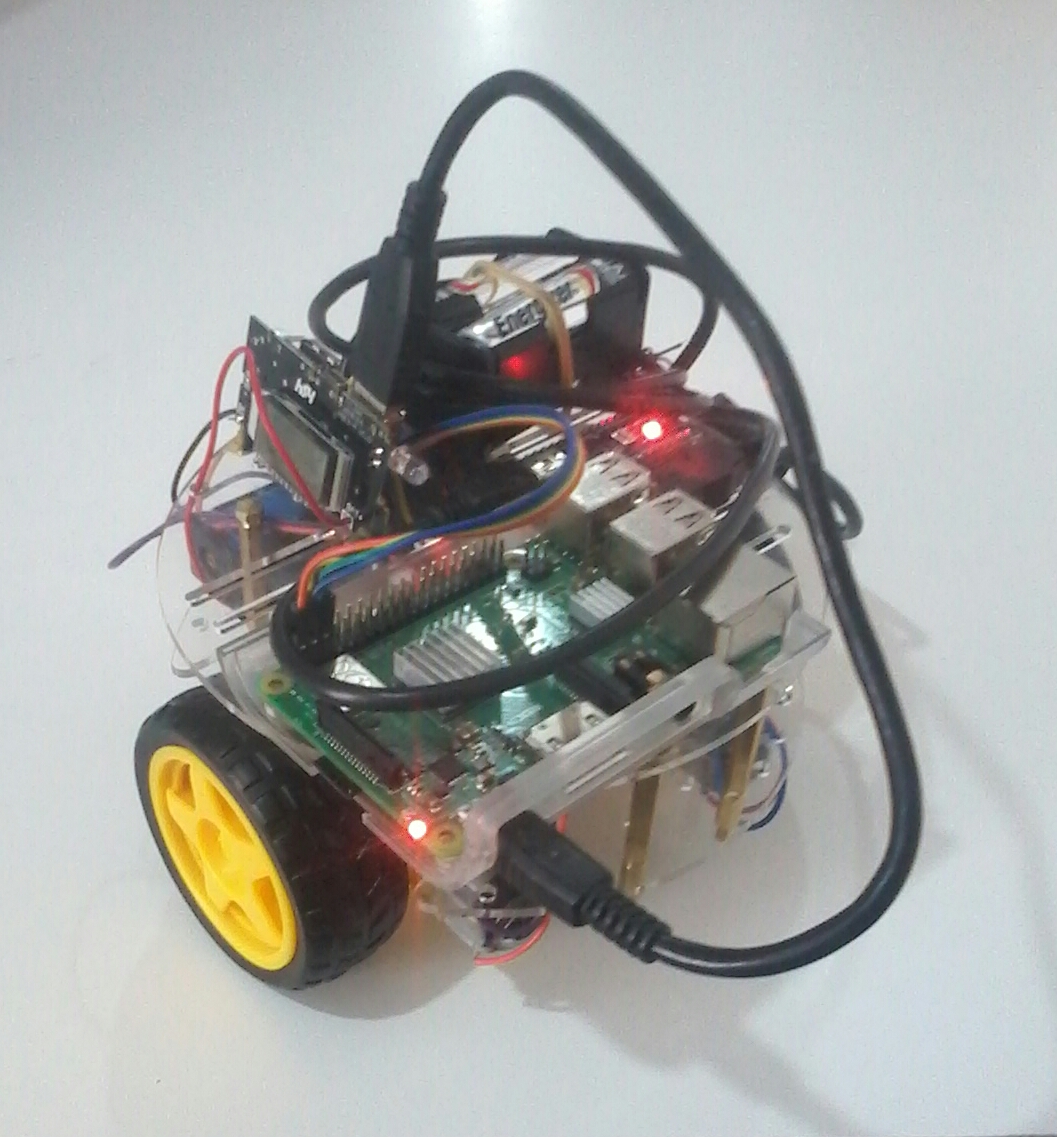}
    \caption{The designed autonomous vehicle}
    \label{fig:foto_robot}
\end{figure}

An image of the complete system can be seen in the figure \ref{fig:foto_robot}

\section{Modeling and Parameter Identification}
\label{sec:modelado}

For the realization of the controller, the positioning system and guidance system, it is necessary to have a model that can predict the response of the system to the control actions used. 

As mentioned in section \ref{sec:introduction}, obtaining the parameters of these models can be a complicated and tedious task. If the user wishes to modify the physical structure of the vehicle in any way, or if the components are degraded by use, the parameters should be updated to reflect these changes in the physical system.

For this reason, the models were made and the following methodologies were designed to obtain their parameters.

\subsection{Dynamic Model}

Following the nomenclature developed in Fossen \cite{Fossen}, the vehicle dynamics respond to the following equations

\begin{equation}
\begin{gathered}
M\dot{\nu} + C(\nu)\nu + D(\nu)\nu + g(\eta) = \tau \\
\dot{\eta} = J(\psi)\nu
\end{gathered}
\label{equ:movement_equation}
\end{equation}

where $eta$ is the vehicle pose (its position and orientation relative to a global coordinate system) $J(\psi)$ is a matrix that relates a vector in a rigidly aligned coordinate system to the global coordinate system, $psi$ is the relative orientation of the vehicle to the global coordinate system, $nu$ is the vehicle speed relative to the vehicle coordinate system, and $u$ are the control actions. 

The matrix $M$ corresponds to the mass of the vehicle, $C$ describes the effects of coriolis and centripetal forces, and $D$ models the effects caused by friction.

As the vehicle can only move in one plane, $\eta$ can be correctly described with a vector $\in R^{3}$, being its components $x, y and, \psi$. The matrix $J$ will then be

\begin{equation}
\begin{bmatrix}
cos(\psi) & -sin(\psi) & 0\\ 
sin(\psi) & cos(\psi) & 0\\ 
0 & 0 & 1 
\end{bmatrix}
\label{equ:J}
\end{equation}
 
If we take into account that the rotation axis and the center of mass are located in the same point, it is feasible to approximate $M$ and $D$ as diagonal matrices, obtaining the following expressions

\begin{equation}
M =
\begin{bmatrix}
m & 0 & 0\\ 
0 & m & 0\\ 
0 & 0 & I 
\end{bmatrix}
\label{equ:M}
\end{equation}

\begin{equation}
C(\nu) =
\begin{bmatrix}
0 & 0 & -m\nu_{y}\\ 
0 & 0 &  m\nu_{x}\\ 
m\nu_{y} & -m\nu_{x} & 0 
\end{bmatrix}
\label{equ:C}
\end{equation}

\begin{equation}
D(\nu) =
\begin{bmatrix}
-d_{l_{x}} - d_{c_{x}} \left\| \nu_{x} \right\| & 0 & 0\\ 
0 & -d_{l_{y}} - d_{c_{y}} \left\| \nu_{y} \right\| & 0\\ 
0 & 0 & -d_{l_{\psi}} - d_{c_{\psi}} \left\| \nu_{\psi} \right\| 
\end{bmatrix}
\label{equ:D}
\end{equation}

Thus, the dynamics can be expressed as

\begin{equation}
\begin{gathered}
\dot{\nu} = \Phi\nu + M^{-1}\tau  \\
\dot{\eta} = J(\psi)\nu
\end{gathered}
\label{equ:dinamic_model}
\end{equation}

$\Phi$ beeing

\begin{equation}
\Phi =
\begin{bmatrix}
\frac{d_{l_{x}} - d_{c_{x}}\left\| \nu_{x} \right\|}{m} & 0 & \nu_{y}\\ 
0 & \frac{d_{l_{y}} - d_{c_{y}}\left\| \nu_{y} \right\|}{m} & -\nu_{x} \\ 
\frac{-m\nu_{y}}{I} & \frac{m\nu_{x}}{I} & \frac{d_{l_{\psi}} - d_{c_{\psi}}\left\| \nu_{\psi} \right\|}{I}
\end{bmatrix}
\label{equ:Psi}
\end{equation}

The torque $\tau$ applied to the system can be characterized as $\tau(u) = Tu$, where $T \in R^{3x3}$. As the dynamics of the motors are much faster than the dynamics of the whole system, this is a good approximation. On the other hand, torque modeling with the $T$ matrix allows us to take into account effects such as differences in the traction performed by each motor, such as non-alignments that cause unmodelled turns, differences in the applied torques, stresses to the motors, etc.

With the model obtained in \ref{equ:dinamic_model} the unknown parameters are: the linear friction coefficients $d_{l_{x}}$ $d_{l_{y}}$ $d_{l_{\psi}}$, the quadratic friction coefficients $d_{c_{x}}$ $d_{c_{y}}$ $d_{c_{\psi}}$ and the matrix $T$. The mass $m$ and the moment of inertia $I$ can be easily measured a-priori, using a balance and the known relationship for the moment of inertia for cylindrical shaped systems $I=\frac{mR^{2}}{1}$, where $R$ is the radius of the cylinder.





In this case, only the $V$ matrix is unknown and must be estimated.

\subsection{Probabilistic Model}

To implement position estimators such as those developed with extended kalman filters \cite{Bao_and_Li_2011}, it is necessary to know the covariance matrices $Q$ and $R$, which model the uncertainty in the sensor measurements and in the model predictions, respectively.

Although it would be possible to obtain the $Q$ matrix from the sensor's data sheet, as the sensors are low cost, many times that information is either not available, of low quality or the common degradation in these types of sensors makes it variable in time, so it is desirable to have a methodology to obtain this data directly from the sensor used. This methodology is described in the following section.

\subsection{Parameter Identification}
\label{Identificacion_de_parametros}

For the estimation of the parameters, a system of minimization of a cost function was implemented:

\begin{equation}
C(p) = \sum_{t} \left\| s(t) - h_{p}(t) \right\|_{h}
\label{equ:cost_function}
\end{equation}

where $s(t)$ are the measurements made by the on-board sensors at each instant of time, $h_{p}(t)$ is a function that predicts the response of the sensors, from the current state of the vehicle, $p$ is the vector of parameters, relative to the specific model and $\left\| \right\|_{h}$ is Huber's norm.





To find the parameters $p$, a nonlinear Gauss Newton optimization approach was used.

The $Q$ and $R$ matrices are obtained using the definition of covariance

\begin{equation}
E((a - E(a))(b - E(b)))
\label{equ:covarianza}
\end{equation}

In order to obtain the expected values, we performed a simulation. The expected values are the real values provided to simulate the system, which then are correlated against the values estimated by the the model. This way be can easily calculate the predictive precision of the parameters obtained by our Gauss-Newton optimization approach.


\section{Implementation of the Subsystems}

\subsection{Positioning system using EKF}

The positioning system was implemented by means of an extended kalman filter. This position estimation system was chosen because it is not computationally demanding and can be perfectly implemented in the embedded system used. The equations governing this filter follow the nomenclature developed in \cite{Thrun_et_al_2005} and are as follows 

\begin{equation}
\begin{gathered}
Prediction: \\
\overline{\mu_{t}} = g(u, \mu_{t-1}) \\
\overline{\Sigma_{t}} = G_{t}\Sigma_{t-1}G_{t}^{T} + R_{t} \\
K_{t} = \overline{\Sigma_{t}}H_{t}^{T}(H_{t}\overline{\Sigma_{t}}H_{t}^{T} + Q_{t})^{-1} \\
Measurement:\\
\mu_{t} = \overline{\mu_{t}} + K_{t}(z_{t} - h(\overline{\mu})) \\
\Sigma_{t} = (I - K_{t}H_{t})\overline{\Sigma_{t}}
\end{gathered}
\label{equ:EKF}
\end{equation}

The state prediction function $g(u, \mu_{t-1})$, depends on the control action and the pose estimated in the previous instant, and responds to some of the models developed in the section \ref{sec:modelado}.

The state in this case was defined taking into account the sensors on board, and corresponds to the speed and acceleration of the vehicle $\mu = [\nu, \dot{\nu}]$. It was defined in this way because the on-board sensors can only directly measure $\dot{\nu_{x}}$, $\dot{\nu_{y}}$ and $\nu_{psi}$.











For the dynamic model described above, $g(u, \mu_{t-1})$ is 

\begin{equation}
g(u, \mu_{t-1}) =
\begin{bmatrix}
\nu_{t-1} + \dot{\nu}_{t-1}\delta_{t} \\ 
\Phi \nu + M^{-1}\tau \\ 
\end{bmatrix}
\end{equation}

This way, $G_{t}$ is

\begin{equation}
\begin{bmatrix}
1 & 0 & 0 & \delta_{t} & 0 & 0 \\ 
0 & 1 & 0 & 0 & \delta_{t} & 0 \\ 
0 & 0 & 1 & 0 & 0 & \delta_{t} \\ 
\frac{d_{l_{x}} + 2d_{c_{x}}\left\| \nu_{x} \right\| }{m} & \nu_{\psi} & \nu_{y} & 0 & 0 & 0\\ 
-\nu_{\psi} & \frac{d_{l_{y}} + 2d_{c_{y}}\left\| \nu_{y} \right\|}{m} & -\nu_{x} & 0 & 0 & 0 \\ 
\frac{m\nu_{y}}{I} & -\frac{m\nu_{x}}{I} & \frac{d_{l_{\psi}} + 2d_{c_{\psi}}\left\| \nu_{\psi} \right\|}{I} & 0 & 0 & 0
\end{bmatrix}
\end{equation}


The function $h(\overline{\mu})$ is

\begin{equation}
h(\overline{\mu(t)}) = [\dot{\nu}_{x}(t-1), \dot{\nu}_{y}(t-1), \nu_{\psi}(t-1)]
\label{equ:h}
\end{equation}

and H 

\begin{equation}
\begin{bmatrix}
0 & 0 & 0 & 1 & 0 & 0 \\ 
0 & 0 & 0 & 0 & 1 & 0 \\ 
0 & 0 & 1 & 0 & 0 & 0 \\ 
\end{bmatrix}
\label{equ:H}
\end{equation}

This implementation 
rely heavily on the estimation of the model parameters. In turn, it is necessary to use the vehicle model for position estimation, since the on-board sensor only gives direct measurements of $\dot{\nu_{x}}$, $\dot{\nu_{y}}$ and $\nu_{\psi}$, so the only way to avoid integration noise is by integrating the sensor measurements with the model prediction.

\subsection{Base Controller}

The implementation of the controller system was done by using a classic PD controller

\begin{equation}
\begin{gathered}
e = J^{-1}(\eta_{r}(t) - \eta_{e}(t)) \\
u_{x} = \alpha e_{x} \\
u_{y} = 0 \\
u_{\psi} = \beta e_{\psi} + \gamma dot(e_{\psi})
\end{gathered}
\end{equation}

The parameters $\alpha$, $\beta$ and $\gamma$ were chosen to minimize the cost function:

\begin{equation}
C(p) = \sum_{t} \left\| \eta_{ref}(t) - \eta_{e}(t) \right\|_{h}
\label{equ:controller_cost_function}
\end{equation}

Where $\eta_{e}$ is obtained by simulating the system with the dynamic parameters that model the real vehicle.

\subsection{Guidance System}

A commonly used guidance system in literature are the ones named \emph{Teach and Repeat}. These systems are based on teaching a trajectory to the autonomous system once, and then it repeats the trajectory autonomously. The possible applications of these types of systems are the movement of supplies or extraction of materials, which require navigating the same path several times. This type of system has very good results in the literature \cite{Furgale_and_Barfoot_2010}.

A system \emph{Teach and Repeat} was implemented guiding the vehicle remotely using a computer through commands sent by ssh connection, and then the vehicle travels the same path indefinitely.

\section{Case Study I: Simulated Vehicle}

As a first evaluation method, a simulation of a vehicle was implemented, following the equations described in the section \ref{sec:modelado}. In order to make the simulation similar to subsequent real experiments, the following parameters were chosen: $m = 1.47 kg$, $I = 810.44$, $d_{l_{x}}$ $d_{l_{y}}$ $d_{l_{\psi}} = [-7.0, -7.0, -500.553]$, $d_{c_{x}}$ $d_{c_{y}}$ $d_{c_{\psi}} = [-3.5, -3.5, -250.0]$ and $T = diag(1.0, 1.0, 29.99)$.


\subsection{Parameters Estimation}

For parameters estimation, the simulated vehicle is provided with a control action consisting of a persistent exciting signal. The data obtained by the sensor was saved in a file for subsequent usage.








\subsubsection{Dynamic Parameters}

The following parameters were obtained for the dynamic model

\begin{equation}
\begin{gathered}
dl = [-7.511 -6.4657 -411.7147] \\
dc = [3.4398 -4.6451,-313.1575] \\
T = 
\begin{matrix}
0.98720 & 0.0 & 0.017941 \\
-0.00023045 & 0.0 & -0.0000411\\
-1.244 & 0.0 & 29.29
\end{matrix}
\end{gathered}
\end{equation}

The estimated $Q$ and $R$ matrices were

\begin{equation}
Q =
\begin{bmatrix}
2.852 & -0.006 & -0.018 \\
-0.006 & 0.0  & 0.0 \\
-0.018 & 0.0 & 0.008
\end{bmatrix}
\label{equ:QV_simulado}
\end{equation}

\begin{equation}
R =
\begin{bmatrix}
2.129 & -0.004 & -0.021 & 0.882 & 0.005 & -0.05 \\
-0.004 & 0.0 &  0.0 & -0.003 & 0.0 &  0.005 \\
-0.021 &  0.0 & 0.006 &  0.006 & 0.0 & 0.115 \\
0.882 & -0.003 &  0.006 &  7.099 & -0.032 &  0.357 \\
0.005 & 0.0 & 0.0 & -0.032 & 0.0 & 0.001 \\
-0.058 & 0.005 & 0.115 & 0.357 & 0.001 & 5.064
\end{bmatrix}
\label{equ:RD_simulado}
\end{equation}

The parameters with which the dynamic simulation was carried out could be estimated correctly.

Note that, since we do not enter a control signal in the direction $y$, the parameters related to this direction are not observable, so the whole second row of the $Q$ matrix is zero.

It can be seen that in the estimation of the covariance matrix of the $R$ model, regarding the estimation of the uncertainty in the acceleration estimates, smaller uncertainties were obtained than those obtained in the kinematic model. This is clearly a product of the fact that the model better represents the underlying process, so the predictions are more accurate. As far as velocity estimation is concerned, the kinematic model can estimate it more accurately. This is probably due to the fact that the simulated vehicle has very fast dynamics, so a kinematic model can represent it correctly.

\begin{figure}[h]
    \centering
    \includegraphics[width=1.0\linewidth]{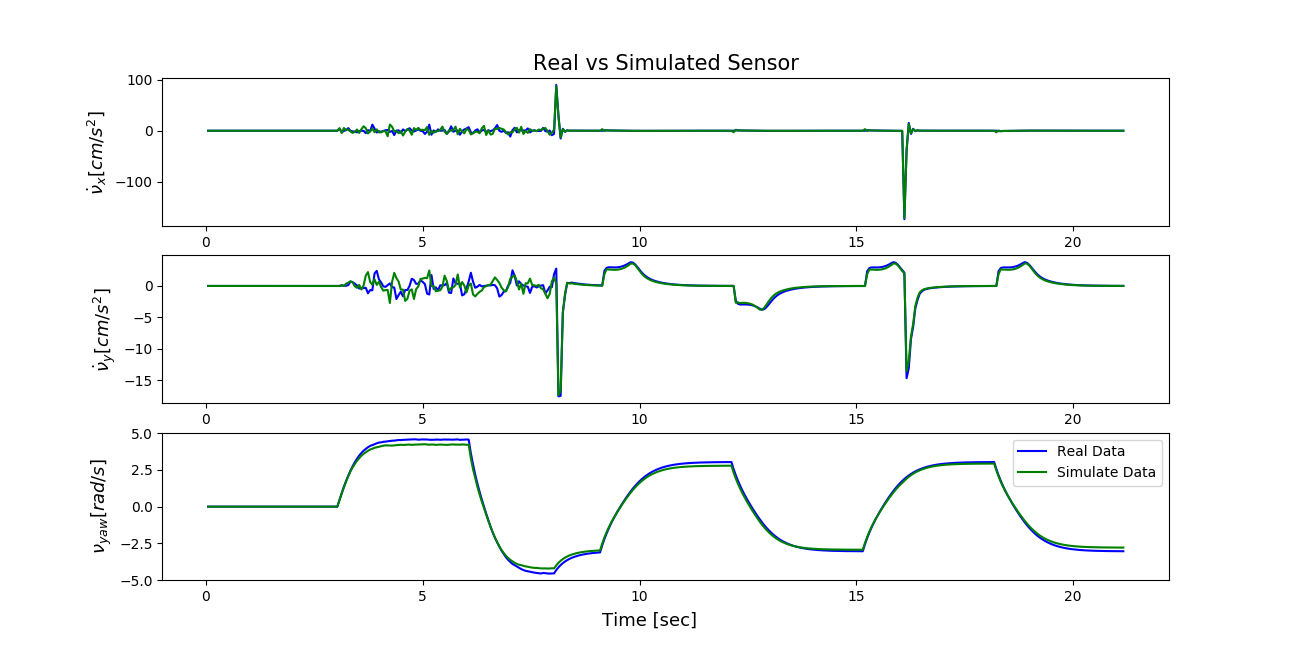}
    \caption{Comparison of sensor measurements during parameter estimation The measurements made by the sensor are shown in blue. In green, the same measurements estimated by the dynamic model of the vehicle}
    \label{fig:acceleracion_simulacion}
\end{figure}

The figure \ref{fig:acceleracion_simulacion}, shows the measurements made by the sensors together with the estimation of them coming from the model. Again, an important correlation is observed, which corresponds to a correctly estimated model.

\section{Case Study II: Real Vehicle}

In this case, data obtained by testing with the vehicle described in the section \ref{sec:disenio_vehiculo} was used.

\subsection{Parameter Estimation}

The vehicle was measured, before making the estimates, both its mass and the radius of the circular platform. These parameters have values of $0.61 kg$ and $0.07 m$ respectively. The value of the moment of inertia on the axis $z$ was calculated using the ratio $\frac{m}{R^{2}}$, with a value of $669.12$.








\subsubsection{Dynamic Parameters}

The estimated dynamic parameters for the vehicle are

\begin{equation}
\begin{gathered}
dl = [-11.9173 -0.1218 \\
T = 
\begin{matrix}
0.6984 & 0.0 & 0.0 \\
0.0 & 0.0 & 0.0 \\
0.0 & 0.0 & 0.000965
\end{matrix}
\end{gathered}
\end{equation}

\begin{equation}
Q=
\begin{bmatrix}
3.497 & -0.272 & 0.046 \\
-0.272 & 1.623 & -0.018 \\
0.046 & -0.018 & 0.038
\end{bmatrix}
\label{equ:Q_real}
\end{equation}
  
\begin{equation}
R=
\begin{bmatrix}
5.928 & 0.014 & 0.020 & 1.592 & -0.151 & 0.361 \\
0.014 & 0.011 & -0.000 & 0.058 & -0.061 & 0.038 \\
0.020 & -0.000 & 0.032 & -0.015 & -0.022 & -0.156 \\
1.592 & 0.058 & -0.015 & 55.163 & 0.163 & 7.670 \\
-0.151 & -0.061 & -0.022 & 0.163 & 1.00 & -0.380 \\
0.361 & 0.038 & -0.156 & 7.670 & -0.380 & 7.957
\end{bmatrix}
\label{equ:R_real_dinamica}
\end{equation}

As expected, using the dynamic model improved the quality of the acceleration estimation made by the model, as is accused in its covariance matrix. Again, we observe the phenomenon by which the speed estimation made by the kinematic model is considerably better than that made by the dynamic model. As mentioned above, this is probably due to the low dynamics present in the real vehicle.


\subsection{Evaluation}

\begin{figure}[h]
   \centering
    \includegraphics[width=1.0\linewidth]{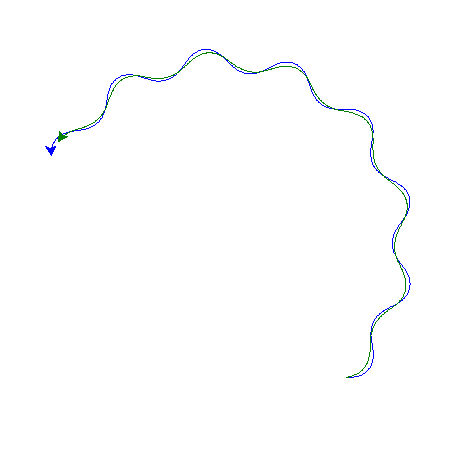}
    \caption{Trajectories obtained during an evaluation of the actual vehicle. In blue is the reference path, provided by the Guidance system in the learning phase. Green is the path estimated by the Kalman filter, using the dynamic model}
    \label{fig:evaluacion_real}
\end{figure}

Finally, an evaluation of the entire system was performed on the actual vehicle. The guidance system was set up so that, in the learning phase, it would make a sinuous circular path. The Kalman filter was used with both the dynamic and the kinematic model, showing similar results. In the figure \ref{fig:evaluacion_real} the trajectory made using the Kalman filter with the dynamic model is shown. Although there are no real data to verify the accuracy of the positioning system when using the real vehicle, it can be mentioned that through visual assessments it is possible to verify that the trajectory was correctly estimated.

\section{Conclutions}

The present work consisted in designing a small, very low-cost autonomous vehicle for use by researchers as teachers. Parameter estimation methodologies were designed, which are necessary when low-cost components are used, for which adequate information is not available, and whose response changes over time due to the degradation suffered by its components. Parameter estimation was evaluated both in simulation and in a real vehicle, and it was found that the physical parameters could be correctly recovered. At the same time, basic implementations of each of the subsystems necessary for the implementation of an autonomous vehicle were carried out. These implementations took into account the low quality of the sensors, being considered the estimated uncertainties in the measurements. The performance of the subsystems was evaluated both in the real vehicle and in simulation. Although the system has an integrative drift in the position estimation, inherent in this type of positioning systems that use intrareceptive sensors, it was possible to obtain accurate estimates of both speed and acceleration. It is the authors' opinion that this work makes an important contribution to the future development of autonomous vehicles, bringing to the community a low-cost system that can be used in a simple and low-cost way.

\end{document}